\documentclass[]{raa}

\usepackage{graphicx,times}

\begin{document}

   \title{Studies of a possible new Herbig Ae/Be star in the open cluster NGC 7380}

\volnopage{{\bf 2012} Vol.{\bf 12} No.{\bf 2}, 167--176}
\setcounter{page}{1} 

   \author{Blesson Mathew
               \inst{1} 
   \and D. P. K. Banerjee
                \inst{1} 
   \and N. M. Ashok
                \inst{1} 
   \and Annapurni Subramaniam
                \inst{2} 
   \and B. Bhavya 
                \inst{2,3} 
   \and Vishal Joshi 
                \inst{1}
          }

   \institute{Astronomy and Astrophysics Division, Physical Research Laboratory,
  Navrangapura, Ahmedabad - 380 009, Gujarat, India; {\it blesson@prl.res.in}\\
         \and
             Indian Institute of Astrophysics, Bangalore - 560 034, India\\
         \and
             Cochin University of Science and Technology, Cochin, India\\
             }

   \date{Received 2011 August 19; accepted 2011 September 26}

\abstract{ We present a study of the star 
2MASS J22472238+5801214 with the aim of identifying its
true nature which has hitherto been uncertain. 
This object, which is a member of the young cluster NGC 7380,
has been variously proposed to be a Be star, a D-type symbiotic and a
Herbig Ae/Be star in separate studies.
Here we present  optical spectroscopy, near-IR photometry and narrow band
H$\alpha$ imaging of the nebulosity in its 
environment. Analysis of all these results, including the spectral energy 
distribution constructed from available data, 
strongly indicate the source to be a Herbig Ae/Be star. 
The star is found to be accompanied by a nebulosity with an interesting
structure. A bow shock shaped structure, similar to a cometary nebula, 
is seen very close to the star with its apex oriented towards the photoionizing 
source of this region (i.e. the star DH Cep). An interesting 
spectroscopic finding, from the forbidden [S{\sc ii}] 6716,
6731 \AA~and [O{\sc i}] 6300 \AA~lines, is the  detection of a blue-shifted high
velocity outflow (200 $\pm$ 50 km s$^{-1}$) from the star.
}

     \authorrunning{B. Mathew et al.}
     \titlerunning{Herbig Ae/Be Star in NGC 7380}

     \maketitle

\section{INTRODUCTION}
\label{sect:intro}

2MASS J22472238+5801214 was identified as a Be
star (category 4B) in the H$\alpha$ emission-line star survey by 
Kohoutek \& Wehmeyer (1997) who 
found strong H$\alpha$ emission in the spectra superposed on a moderate continuum.
On the other hand Corradi et al. (2008) identified this candidate as a D-type symbiotic
binary from the IPHAS H$\alpha$ emission-line survey. This
classification was based on the observed near-IR and H$\alpha$ excess,
from the location in ($r-i$) vs ($r$--H$\alpha$) and ($J-H$) vs
($H-K_s$) color-color diagrams. Symbiotic systems are interacting binaries
with a white dwarf (WD), a cool giant and an emitting nebula, created by the
photoionizing flux from the WD and collision of the winds (Angeloni et
al. 2007). Symbiotic D-type candidates are separated from
S-type candidates based on the continuum excess in 1--4 $\mu$m 
spectral region (Webster \& Allen 1975). 
If the companion belongs to an F/G giant class rather than a Mira variable one,
the system is designated as D' type (Allen 1982).
The spectra of symbiotic binaries are characterized by the presence of high
ionization lines and Raman scattered O{\sc vi} emission lines in addition to
low ionization metallic absorption lines and molecular bands.

Apart from the Be classification of Kohoutek \& Wehmeyer (1997) and the D-type
symbiotic classification of Corradi et al. (2008), the possibility of this object
belonging to the Herbig Ae/Be (HAeBe) category was also suggested by Mathew et al. (2008).
This star was detected in emission in the young cluster NGC 7380 during
the survey of emission-line stars in young open clusters and cataloged as NGC 7380(4).

In all forthcoming text we use the designation 2MASS J22472238+5801214 or NGC
7380(4) equivalently. The object's classification as HAeBe was discussed by
Mathew et al.(2010) based on the following characteristics.
The object showed near-IR excess of $\sim$1 mag
in extinction corrected ($J-H$) versus ($H-K$) color-color diagram. The star was
also found to be located in the position occupied by HAeBe stars in the H$\alpha$ equivalent
width (EW) versus ($H-K$) diagram, which is conventionally used to separate
Classical Be (CBe) and HAeBe stars.

It is thus seen that there is some uncertainty about the true nature of the
object which should desirably be resolved. We attempt to do this by undertaking an
in-depth analysis of the object proper, as also its environment, by using photo-spectroscopic
and imaging data. We believe that we are able to make a secure classification of
the object's nature in the present work. In addition it is also shown that the source
is fairly interesting, by virtue of being associated with a high velocity
outflow, and worthy of further studies.

A few words on the physical environment of the target object in the present
study may be appropriate. Massey et al. (1995) identified
this star as a member of the young open cluster ($\sim$2 Myr) NGC 7380 (star
no: 2249) with visual magnitude $m_V$ = 14.72, color excess $E(B-V)$
= 0.64 and distance 3.6 kpc. The star is located away from
the cluster center and associated with pre-main sequence stars.
The star is less than
0.25 Myr from a pre-main sequence (PMS) isochrone fitting in the $V$ versus ($B-V$) 
color magnitude diagram (Mathew et al. 2010). The target NGC 7380(4) 
is associated with a relatively large ($\theta$ = 25') and evolved H{\sc ii} region
Sharpless 142 (S142; Roy \& Joncas 1985). The main source of ionization is an O6
spectroscopic binary DH Cep, which is also a member of the cluster NGC 7380.
The region is quite complex, showing association with an H{\sc i} cloud and
molecular cloud NGC 7380E (see fig. 1 in Chavarria-K. et al. 1994).

\section{OBSERVATIONS}
\label{sect:Obs}

The spectroscopic observations were done using the HFOSC (Himalayan Faint Object 
Spectrograph Camera)
available with the 2.0m Himalayan Chandra Telescope (HCT), operated by the Indian
Institute of Astrophysics, India.
The CCD used for imaging was a 2 K $\times$ 4 K CCD,
where the central 500 $\times$ 3500 pixels were used for spectroscopy. The
pixel size was 15 $\mu$m
with an image scale of 0.297 arcsec/pixel. The spectra were taken using a 
Grism 7 (3800--6800 \AA) and 167 $\mu$m slit combination in the blue
region which gave an effective
resolution of 10 \AA~near the H$\beta$ line. 
The spectra in the red region were taken using a Grism 8
(5500--9000 \AA) and 167 $\mu$m slit setup, which gave an effective
resolution of 7 \AA~near the H$\alpha$ line. The spectra were found to
have good signal to noise ratio ($\ge$ 100). The HFOSC was also
used in imaging mode to obtain a broad band (6300--6740 \AA) H$\alpha$ image of the
source and its environment. 

$JHK$ photometric observations of the object were made from
Mt. Abu Infrared Observatory on 2010 October 21 using the 256$\times$256 NICMOS3
imager-spectrograph. The procedure for the near-IR photometric observations and the subsequent
reduction and analysis of data followed a standard procedure described e.g
in Banerjee \& Ashok (2002). 
All spectroscopic and photometric data  were reduced and analyzed using IRAF tasks.
A consolidated log of the observations is given in Table 1.

%table 1
\begin{table*}
\caption{Journal of Observations}
 \centering
\begin{tabular}{lccc}
\hline
Object & Date of Observation & Mode of Obs. & Specifics \\
\hline
NGC7380(4) & 2010-12-05 & H$\alpha$ imaging      & exp. 60s, field 10$\arcmin$$\times$10$\arcmin$, HFOSC, 2.0 m HCT\\
            & 2010-12-05 & H$\alpha$ imaging      & exp. 40s, field 2$\arcmin$$\times$2$\arcmin$, HFOSC, 2.0 m HCT\\
            & 2010-10-05 & Spectroscopy & Grism 7/167l, exp. 1200s, HFOSC, 2.0 m HCT \\
            & 2010-10-05 & Spectroscopy & Grism 8/167l, exp. 1200s, HFOSC, 2.0 m HCT \\
            & 2010-10-21 & JHK photometry & NICMOS3, 1.2 m Mt Abu \\
Nebulosity  & 2010-12-05 & Spectroscopy & Grism 8/167l, exp. 2400s, HFOSC, 2.0 m HCT \\
\hline
\end{tabular}
\end{table*}

%figure 1
\begin{figure*}
\centering
\includegraphics[bb=50 275 550 500, width=14cm, clip]{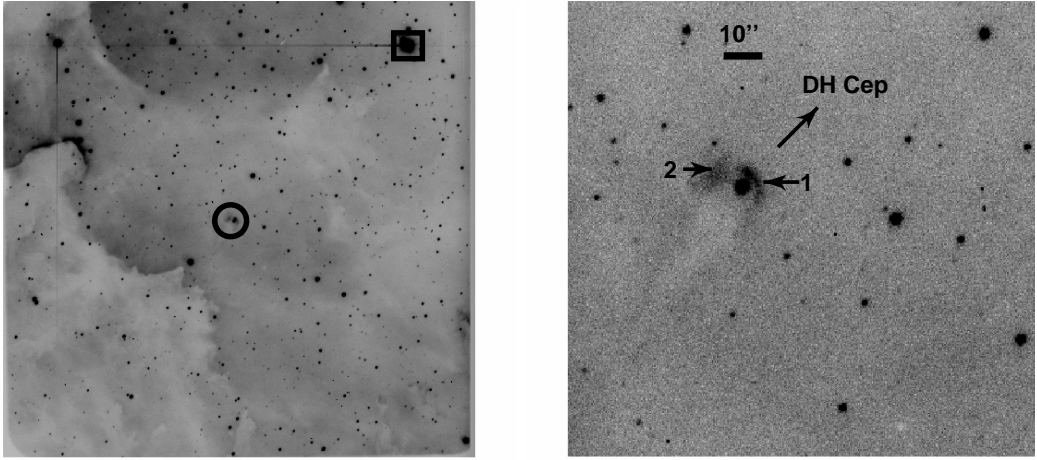}
\caption{Left panel shows a 10$\arcmin$ $\times$ 10$\arcmin$ field around the
object of interest obtained with an H$\alpha$ broad band filter. NGC 7380(4)
and  DH Cep are shown surrounded by a circle and square respectively. The
right panel shows a zoomed image (2$\arcmin$ $\times$ 2$\arcmin$)
of the nebulosity around NGC 7380(4). In both
panels, north is to the top and east to the left.
More details are given in Section 3.1.}
\end{figure*}

\section{RESULTS}

\subsection{H$\alpha$ Imaging: A Nebulosity Around the Object}

The H$\alpha$ image of the region and an enlarged section around the star are
shown in the left and right panels of Figure 1 respectively. A nebulosity is
clearly seen around the object
whose principal features consist of a diffuse patch (feature 2) and a
bow-shock shaped structure (feature 1) very close to the star.
The bow shaped structure looks like a cometary globule (cometary nebula) with the apex,
as expected in these objects, oriented towards the photoionizing 
source which in this particular case is DH Cep.
Cometary globules are potential sites of induced star formation due to
compression by ionization or shock fronts, created by the influx of
UV radiation from the massive exciting star. 
Ikeda et al. (2008) found six H$\alpha$ emission stars near the tip of the
cometary globule BRC 37, which are formed due to the sequential star formation
triggered by O-type stars HD 206267 and HD 206183. 
Sugitani et al. (1991) cataloged forty four bright rimmed clouds with IRAS
point sources, which are
possible candidates for star formation by radiation-driven implosion.
Our candidate was not listed in the catalog
even though S142 was identified, which is seen as a bright rim to the
left of the object in the 10$\arcmin$ $\times$ 10$\arcmin$ field (Fig. 1).

Negueruela et al. (2007) studied triggered star formation in NGC 1893,
which is similar to NGC 7380 in terms of age and star formation activity.
From H$\alpha$ imaging and slitless spectroscopy they identified a Herbig Be
star S1R2N35 in the immediate vicinity of cometary nebula Sim 130 
(a striking image of this cometary nebula is shown in the above work). 
Also one can see bow shaped structure and nebulosity associated with this
region which are triggered by nearby massive stars.
This shows that the presence of a Herbig Be star in the
vicinity of a cometary globule is possible and supports 
an HAeBe classification for NGC 7380(4).

%figure 2
\begin{figure*}
\centering
\includegraphics[width=\textwidth]{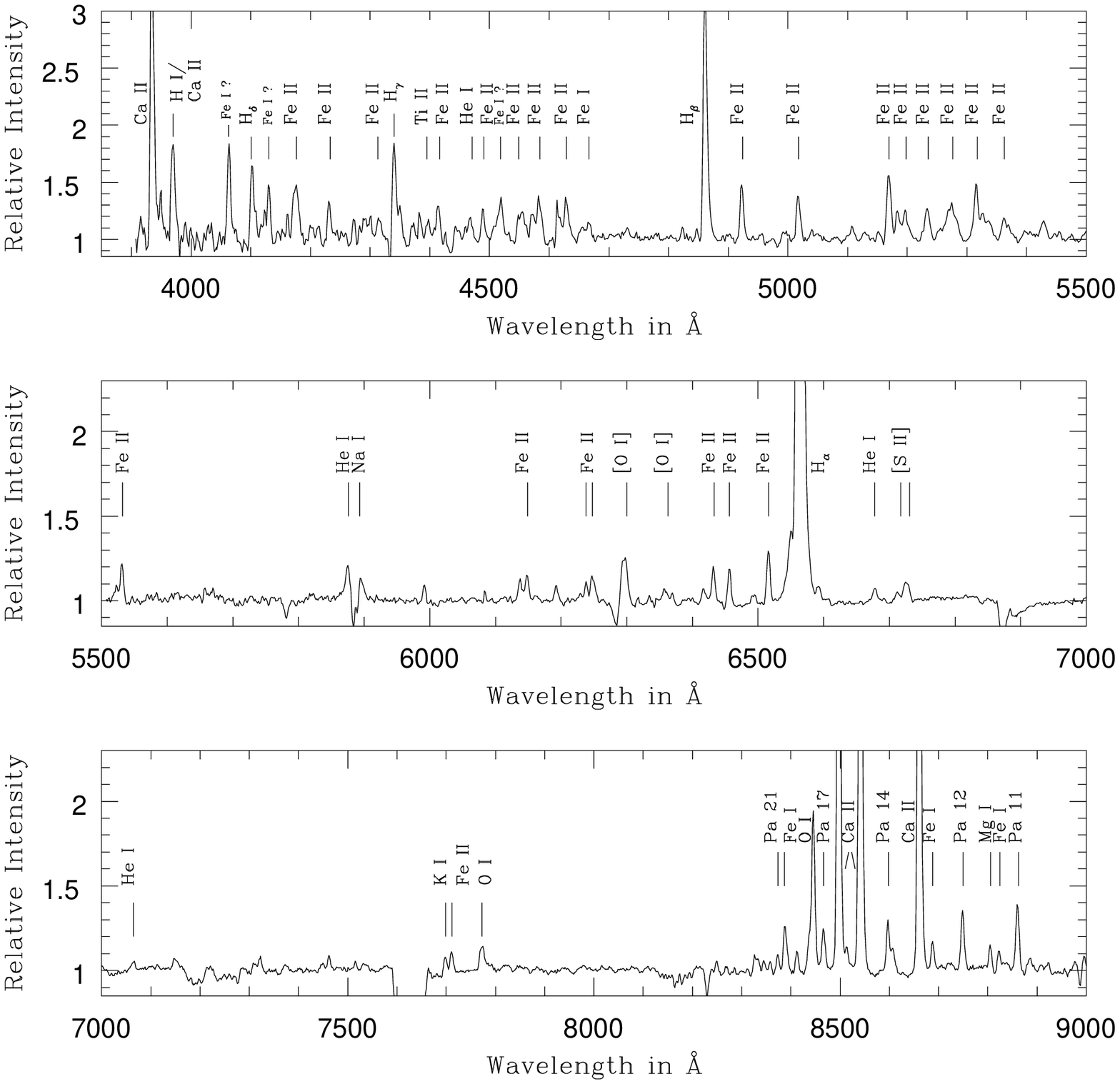}
\caption{Optical spectrum of NGC 7380 between 4000--9000 \AA~taken on 
2010-10-05. The prominent lines are identified.}
\end{figure*}

\subsection{Spectroscopy of the Source}

The optical spectrum of NGC 7380(4) is presented in Figure  2. 
All the lines are seen in emission and absorption features, if any, are not prominent.
Hydrogen lines of the Balmer and Paschen series are all in emission;  it may be noted that the
higher order lines of these series  are usually seen in absorption in the
spectrum of Be stars.
The H$\alpha$ line is the most intense (EW $\sim$ $-$100 \AA) in the spectrum with
broad  wings extending from 6530 to 6595 \AA. The other prominent lines seen
are due to Ca{\sc ii}, neutral lines of Na{\sc i} and K{\sc i}, permitted and
forbidden lines of  O{\sc i}, [S{\sc ii}], a few lines of He{\sc i} and a large
number of lines from Fe{\sc ii}. Line identification is largely based on the
detailed list of lines typically seen in the spectra of HAeBe stars presented
in Hern{\'a}ndez et al. (2004). Several weak features in the spectrum of Figure 2
remain unidentified. Comparison of their wavelengths with atomic line lists
suggests that many of them could be due to Fe{\sc i}. However, a secure identification
is difficult to arrive at and for the present study we leave them as
unidentified. 

Table 2 presents the prominent lines seen in the spectra along
with the equivalent widths which have  typical  measurement
errors of around 5 to 10\%. In the case where lines were blended,  we
employed a deblending procedure involving the fitting of multiple gaussians to
the observed profile. The equivalent widths of the individual gaussians were then
estimated. Lines with uncertain identification are marked with a question mark in the 
first column of Table 2. 

An interesting aspect of the spectra is the evidence of a fast outflow as inferred
from the  behavior of the forbidden lines
of [S{\sc ii}]$\lambda\lambda$6716/6731 and [O{\sc i}]$\lambda\lambda$6300/6364.
The presence of these forbidden lines in the spectra of  HAeBe stars,
and in their lower mass counterparts - the classical T Tauri stars (CTTS), 
has long been used to infer
the presence of jets/outflows since such lines arise only in low density
conditions and hence are tracers of low density material 
(Finkenzeller 1985, Corcoran \& Ray 1998, Appenzeller et al. 1984).

Figure 3 shows a magnified section of the spectra around the forbidden lines showing
these lines to be blue shifted by $\sim$4 to 5 \AA~whereas other lines in
the spectrum are seen at their expected wavelengths.
The measured mean blue-shift for the [S{\sc ii}]$\lambda\lambda$6716/6731 lines is 215
$\pm$ 50 km s$^{-1}$ while that for the [O{\sc i}]$\lambda$6300 line
is 176 $\pm$ 50 km s$^{-1}$. Thus there is evidence for the presence of
a high velocity outflow emanating from the star. 
The ratio of the emission strengths of the [S{\sc ii}] doublet (6716/6731) is around 0.42
indicating that the electron density is close to (or greater than) $\sim$ 10$^4$ cm$^{-3}$
if a temperature of 10000K is assumed (Osterbrock \& Ferland 2006, Canto et al. 1980).
Such a value of the electron density 
is slightly on the higher side compared to H{\sc ii} or nebular regions; 
but similar values have been
observed in certain parts of a similar [S{\sc ii}] outflow 
emanating from the HAeBe star LkH$\alpha$233 (Corcoran \& Ray 1998).
It may be noted that the absorption feature seen to the left  of 
the [O{\sc i}]$\lambda$6300 line, giving
it an apparent P-Cygni structure, is actually a Diffuse Interstellar Band.

%figure 3
\begin{figure}
\centering
\includegraphics[width=8cm]{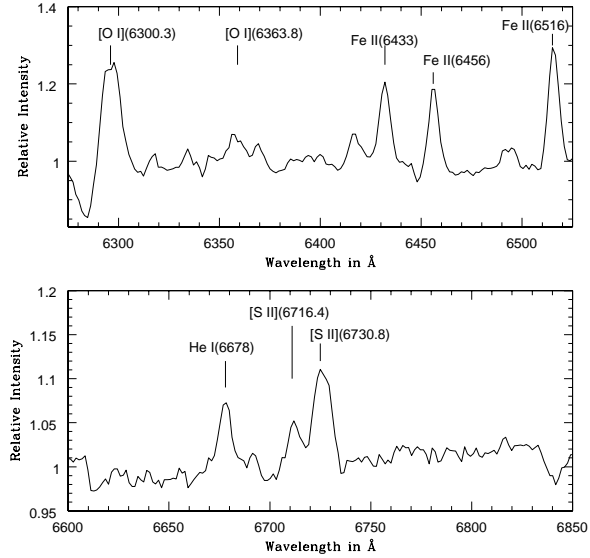}
\caption{[O{\sc i}]$\lambda\lambda$6300, 6364 and 
[S{\sc ii}]$\lambda\lambda$6716, 6731 line profiles observed in the object.}
\end{figure}

 %figure 4
\begin{figure}
\centering
\includegraphics[width=8cm]{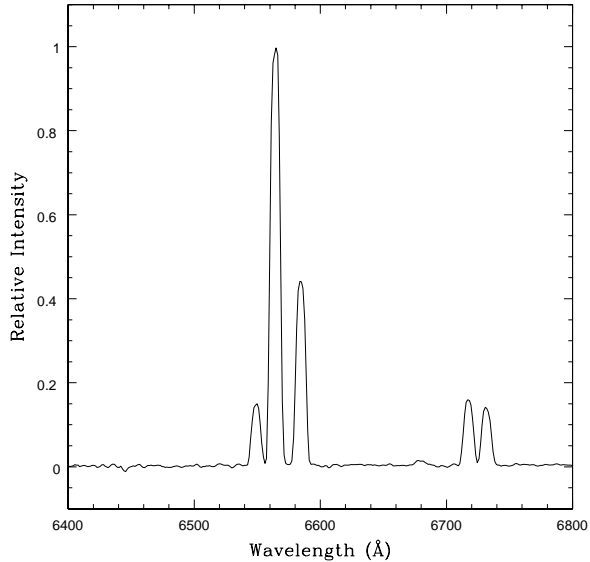}
\caption{Spectrum of the nebulosity feature 2 marked in Fig. 1.
The prominent lines seen are [N{\sc ii}]$\lambda\lambda$6548, 6583;
H$\alpha$ and [S{\sc ii}]$\lambda\lambda$6716, 6731. He{\sc i}$\lambda$6678
is also weakly seen}
\end{figure}

The Ca{\sc ii} triplet (8498 \AA, 8542 \AA, 8662 \AA) lines are blended
with the Paschen lines Pa16, Pa15 and Pa13 respectively.
The contribution of these Paschen lines is estimated by interpolating the
strengths of the isolated adjacent Paschen lines Pa17, Pa14 and Pa12 and 
removed from the Ca{\sc ii} triplet line strengths. From these corrected
equivalent width values,
the relative strength of triplet lines are found
to be in the ratio 1.0 : 0.98 : 0.84. This is vastly different from the expected
strengths of 1 : 9 : 5, which is the ratio of their respective {\it gf}
values. This implies that the Ca triplet lines are subject to large optical depth effects.
It may be noted that the intensity of the 8498 \AA~line  is greater than the 8542 \AA~line,
which is a unique 
characteristic of PMS stars (Hamann \& Persson 1992).

Could NGC 7380(4) be a symbiotic star? Based on the examples
of well-studied and widely accepted symbiotic objects Belczy{\'n}ski et al. (2000) 
adopted the following spectral criteria to classify an object as a
symbiotic star: (i) the presence of absorption features of a late-type giant like TiO,
H$_2$O, CO, CN, or VO bands as well as Ca{\sc i}, Ca{\sc ii}, Fe{\sc i}, or Na{\sc i}
absorption lines (ii) the presence of strong emission lines of H{\sc i} and He{\sc i}
and either emission lines of ions with ionization potential of at least 35 eV
like [O{\sc iii}] or high ionization lines from
[Fe{\sc vii}]$\lambda\lambda$5721, 6086, He{\sc ii}$\lambda\lambda$4686, 5411
and Ca{\sc v}$\lambda$6086 (Corradi \& Giammanco 2010) 
(iii) the presence of Raman scattered 6825 \AA~emission feature.
Schmid (1989) identified Raman scattered O{\sc vi} lines 
(6825 \AA~and 7082 \AA) in the spectra of symbiotic binaries,
which are not observed in other astrophysical objects.
These lines are produced by Raman scattering of
the O{\sc vi}$\lambda\lambda$1032/1038 resonance lines by neutral hydrogen. 
Since none of these criteria are met in the case of NGC 7380(4) it is unlikely to
be a symbiotic star. Further, the attributed association of
the star with a young cluster whose age is 2 Myr indicates it to be a young
object; symbiotic stars are relatively more evolved systems as implied by the
presence of a WD as one of the components.

A spectrum (5500--9000 \AA) of the nebulosity (feature 2 in Fig. 1)  was taken with
the slit positioned along NS (PA = 0$^o$) and an exposure time of 2400 s.
The spectrum is typically nebular with a weak continuum, which barely
registers above the dark counts of the detector, with the prominent lines being
[N{\sc ii}]$\lambda\lambda$6548/6583, H$\alpha$ and
[S{\sc ii}]$\lambda\lambda$6716/6731. This part
of the spectrum is shown in Figure 4. Very few additional lines are seen and
these are He{\sc i}$\lambda\lambda$5876, 6678, [O{\sc i}]$\lambda$6300 and an unidentified line at
7136 \AA~(possibly [Ar{\sc ii}]). It is possible that this nebulosity could be 
partially a reflection nebulosity and partially an ionized region. 
The observed [S{\sc ii}] (6716/6731)
ratio of 1.16 in the nebulosity implies a electron density in the range 
$\sim$100--150 cm$^{-3}$ assuming a
temperature of 10000 K. It is difficult to be certain whether the region is
shock ionized or photo-ionized by the UV flux from DH Cep. 
In shock ionization, low-ionization lines like
[S{\sc ii}]$\lambda\lambda$6716/6731 are much stronger
with respect to H$\alpha$ than in typical photoionized H{\sc ii} regions
(Osterbrock \& Ferland 2006; Hartigan et al. 1994; the H$\alpha$ to S{\sc ii}[6717 + 6731]
ratio can be around unity). For a representative
comparison, Osterbrock \& Ferland (2006) listed line intensities in the Orion nebula
(photoionized) and a shock ionized filament in Cas A. The observed
I(H$\alpha$)/I(6716) ratio is about 90 in the former
and 2.6 in the latter. In our case I(H$\alpha$)/I(6716) has a value of $\sim$6.4,
closer to that expected in a shock-ionized region. 
Thus a part of the [S{\sc ii}] emission seen in the
nebulosity may arise from a  shock. 
However, a deeper study of this region is desirable, to draw firmer conclusions. 

\begin{table*}
\begin{center}
\caption{Emission lines in NGC 7380(4)}
\begin{tabular}{lcccccccccc}
\hline
Element & $\lambda$ & EW & & Element & $\lambda$ & EW & & Element & $\lambda$ & EW \\
        & (\AA~) & (\AA~) & &        & (\AA~) & (\AA~) & &        & (\AA~) & (\AA~) \\
\hline
Ca{\sc ii}K & 3933 & $-$19.1 & & H$\beta$ & 4861 & $-$18.3 & & H$\alpha$ & 6563 & $-$100.1 \\
Ca{\sc ii}/H{\sc i} & 3970 & $-$6.5 & & Fe{\sc ii}(42) & 4924 & $-$3.6 & & He{\sc i} & 6678 & $-$0.7 \\
Fe{\sc i} ? & 4063 & $-$4.1 & & Fe{\sc ii}(42) & 5018 & $-$3.1 & & $[$S{\sc ii}$]$ & 6716 & $-$0.6 \\
H$\delta$ & 4101 & $-$5.4 & & Fe{\sc ii}(42) & 5169 & $-$5.6 & & $[$S{\sc ii}$]$ & 6731 & $-$1.4 \\
Fe{\sc i} ? & 4130 & $-$3.2 & & Fe{\sc ii}(49) & 5198 & $-$3.9 & & He{\sc i} & 7065 & $-$0.4 \\
Fe{\sc ii}(27,28) & 4176 & $-$5.5 & & Fe{\sc ii}(49) & 5235 & $-$2.8 & & Fe{\sc ii} & 7712 & $-$1.0 \\
Fe{\sc ii}(27) & 4233 & $-$1.8 & & Fe{\sc ii}(49) & 5276 & $-$5.4 & & O{\sc i} & 7772 & $-$2.1 \\
Ti{\sc ii}(41) & 4313 & $-$1.2 & & Fe{\sc ii}(48,49) & 5317 & $-$4.2 & & Pa21 & 8374 & $-$0.8 \\
H$\gamma$ & 4340 & $-$6.6 & & Fe{\sc ii}(49) & 5326 & $-$1.5 & & Pa20/Fe{\sc i} & 8387 & $-$2.2 \\
Fe{\sc ii}(27) & 4352 & $-$3.9 & & Fe{\sc ii}(48) & 5338 & $-$2.4 & & Pa19 & 8413 & $-$0.9 \\
He{\sc i}+Fe{\sc ii}(27) & 4385 & $-$1.2 & & Fe{\sc ii}(48) & 5363 & $-$1.7 & & O{\sc i} & 8446 & $-$9.7 \\
Ti{\sc ii}(19) & 4395 & $-$0.9 & & Fe{\sc ii}(55) & 5535 & $-$0.6 & & Pa17 & 8467 & $-$2.4 \\
Fe{\sc ii}(27) & 4417 & $-$1.8 & & He{\sc i} & 5876 & $-$1.6 & & Ca{\sc ii} & 8498 & $-$34.3 \\
He{\sc i} & 4471 & $-$1.9 & & Na{\sc i} & 5890/96 & $-$0.9 & & Ca{\sc ii} & 8542 & $-$33.5 \\
Fe{\sc ii}(37) & 4491 & $-$1.9 & & Fe{\sc ii}(74) & 6149 & $-$0.9 & & Pa14 & 8598 & $-$2.1 \\
Fe{\sc i}/Fe{\sc ii} ? & 4519 & $-$2.6 & & Fe{\sc ii}(74) & 6238 & $-$0.5 & & Ca{\sc ii} & 8662 & $-$29.0 \\
Fe{\sc ii}(38) & 4549 & $-$3.5 & & Fe{\sc ii}(74) & 6248 & $-$1.3 & & Fe{\sc i} & 8688 & $-$2.0 \\
Fe{\sc ii}(37,38) & 4584 & $-$2.6 & & $[$O{\sc i}$]$ & 6300 & $-$3.3 & & Pa12 & 8750 & $-$3.3 \\
Fe{\sc ii}(38) & 4621 & $-$2.6 & & Fe{\sc ii}(40) & 6433 & $-$1.4 & & Mg{\sc i} & 8806 & $-$1.1 \\
Fe{\sc ii}(37) & 4629 & $-$3.6 & & Fe{\sc ii}(74) & 6456 & $-$1.4 & & Fe{\sc i} & 8824 & $-$1.0 \\
Fe{\sc i}(37) & 4667 & $-$1.0 & & Fe{\sc ii}(40) & 6516 & $-$1.9 & & Pa11 & 8862 & $-$4.1 \\
\hline
\end{tabular}
\end{center}
\end{table*}

 %figure 5
\begin{figure}
\centering
\includegraphics[width=8cm]{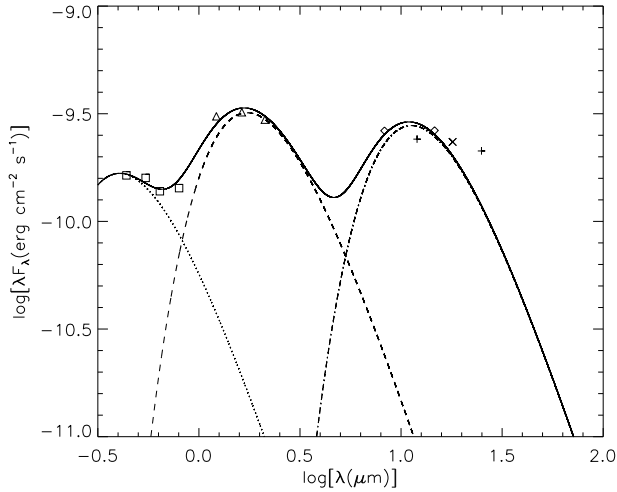}
\caption{Spectral energy distribution of the source is shown using the
  data in Table 3. $BVRI$ points are shown in squares, $JHK_s$ in triangles, MSX
  in diamonds, AKARI in crosses and IRAS in `+' symbols.
Blackbody fits at  temperatures of  9100 K ({\it dotted} line), 
2100 K ({\it dashed} line) and  300 K ({\it dot-dashed} line) are shown
along with their co-added sum which is  shown by a solid line.}
\end{figure}

\subsection{Spectral Energy Distribution}
The photometric data spanning the optical to mid-infrared spectral region are
presented in Table 3. These data are used to construct the spectral energy
distribution (SED), which is shown in Figure 5.
It should be noted that the optical, near-IR and mid-IR observations are done
at different epochs. The two sets of near-IR measurements separated by ten
years do not show noticeable variability. The reddening
corrections were done using relations from Rieke \& Lebofsky (1985) 
with $E(B-V)$ = 0.64 (Massey et al. 1995).
The SED shows a clear IR excess and the IR luminosity is significantly larger
than the optical luminosity. This is a typical characteristic
of HAeBe stars belonging to the Group II class (Hillenbrand et al. 1992).
The SED of Group II objects is interpreted in terms of a spherical envelope
and regarded as the precursors to HAeBe stars with a circumstellar
disk. We have fitted multiple blackbodies to identify different components in the
SED. The multiple blackbody fit suggests the presence of a hot component
with a temperature of $\sim$9100 K and two additional components likely to be associated with
dust, with temperature of $\sim$2100 K and $\sim$300 K respectively.
However, the use of an appropriate  radiative-transfer code like DUSTY 
is necessary, which is beyond the scope of this paper, to properly 
estimate the physical parameters of
the dust envelope surrounding the central star.

%table 3
\begin{table}
\begin{center}
\caption{Available Optical-IR Photometric Measurements}
\begin{tabular}{lccc}
\hline
Source & Wavelength/band & Flux/mag \\
\hline
NOMAD     & $B$               & 15.69  \\
Massey et al. (1995) & $V$    & 14.72  \\
IPHAS     & $R$               & 14.00  \\
          & H$\alpha$       & 12.85 \\
          & $I$               & 12.95  \\
2MASS (Mt. Abu)     & $J$               & 10.80 (10.88)\\
          & $H$               & 9.76 (9.71)  \\
          & $K_s$               & 8.85 (8.75) \\
MSX6C     & 8.28 $\mu$m     & 0.73 Jy\\
          & 14.65 $\mu$m    & 1.29 Jy\\
IRAS      & 12 $\mu$m       & 0.96 Jy\\
          & 25 $\mu$m       & 1.77 Jy\\
AKARI     & 18 $\mu$m       & 1.40 Jy\\
\hline
\end{tabular}
\end{center}
\end{table}

\subsection{HAeBe Nature of the Candidate}

Herbig (1960) classified HAeBe stars on the basis of the following criteria: 
(a) the spectral type is A or earlier, with emission lines, (b) the star lies
in an obscured region, and (c) the star illuminates a fairly bright nebulosity in
its immediate vicinity. 
Waters \& Waelkens (1998) modified the above definition and removed the constraint of an
associated nebulosity  by considering the fact
that isolated HAeBe stars are also seen, which were identified from the IRAS
far-IR all sky survey. Hence they propose the present working
definition of HAeBe stars  as: (a) spectral type A or B with emission lines, (b) infrared (IR)
excess due to hot or cool circumstellar dust or both, and (c) luminosity class III
to V. In the following discussion we have analyzed the merits of NGC 7380(4) as
 an HAeBe candidate.

As explained in Section 3.2, the spectra of NGC 7380(4) show emission
lines. The estimation of spectral type from spectroscopy is not possible since absorption
lines of hydrogen and helium are absent. Thus, if the
spectral class is to be identified even in a very broad sense, we have to take recourse to 
photometric data. Using several stars in the cluster, Massey et al. (1995) 
estimated the distance of the cluster to be 3732 pc and also estimated a mean
reddening to be $E(B-V)$ = 0.64. Using this value of the reddening and 
an apparent magnitude m$_V$ = 14.72 for the object,  
an absolute magnitude of M$_V$ = $-$0.12 was derived. This would correspond to
a B8 -- B9 spectral type if it is of luminosity class V and B9 -- A0 if it is
of luminosity class III (Schmid-Kaler 1982). 
However, there is likely to be a variation in the intra-cluster reddening as
shown by Massey et al. (1995) whose sample of stars showed a variation in
$E(B-V)$ between 0.52 to 0.86. 
Therefore, using the mean value of $E(B-V)$ = 0.64 could lead
to errors in estimating the  absolute magnitude  and hence the
spectral type of the star. 
Thus, the photometric data broadly suggest that NGC 7380(4) is a late B
or early A type star, which is in line with the requirement for it to be an HAeBe star.
From its SED we identified IR excess in this star, which is considered as
a defining property of HAeBe stars. 
The star is also associated with a nebulosity whose 
presence further strengthens the HAeBe classification of the object.
The spectroscopic support for such a classification has already been discussed.

\section{SUMMARY}

We have presented a study of the object 
NGC 7380(4) (equivalently 2MASS J22472238+5801214) whose classification was
hitherto uncertain. The star is shown to satisfy many of the characteristics of
HAeBe stars viz. a similar spectrum, 
association with a star forming region, an SED showing an infra-red excess 
that is expected of this category
of stars, the presence of a surrounding nebulosity and a suggested young age
by virtue of being associated with the young cluster NGC 7380.
It is thus strongly suggested that the object is an HAeBe star rather than a
D-type symbiotic or a Be star. 
We find spectroscopic evidence, based on the forbidden lines of [S{\sc ii}] and [O{\sc i}],
for the interesting presence of a 200 $\pm$ 50 km s$^{-1}$ high velocity outflow originating from
the star. From H$\alpha$ imaging, a nebulosity is clearly seen around the object
whose principal features consist of a diffuse patch (east of the star) and a
bow-shock shaped structure typical of a cometary nebula. 
The apex of this cometary nebula is seen to point 
towards the star DH Cep which is believed to be the hot photoionizing source of this 
region. Such an orientation of the cometary nebula towards the ionizing source
is generally seen in other similar objects. \\

\begin{acknowledgements}
The research work at Physical Research Laboratory is funded by the Department
of Space, Government of India.
We would like to acknowledge the assistance of Pepsi Anto, the staff in Hanle and
those in Mt.Abu during the observations.
\end{acknowledgements}

\label{lastpage}


\begin{thebibliography}{99}

\bibitem[1982]{alle82} Allen, D.~A. 1982, in IAU Colloq. 70: 
The Nature of Symbiotic Stars, Vol.95, eds., M.~Friedjung \& R.~Viotti, 27

\bibitem[2007]{ange07} Angeloni, R., Contini, M., Ciroi, S., Rafanelli, P. 2007,
  \aap, 472, 497

\bibitem[1984]{appe84} Appenzeller, I., Oestreicher, R., Jankovics, I. 1984, \aap, 141, 108

\bibitem[2002]{bane02} Banerjee, D.~P.~K., Ashok, N.~M. 2002, \aap, 395, 161

\bibitem[2000]{belc00} Belczy{\'n}ski K., Miko{\l}ajewska J., Munari U., 
Ivison, R.~J., Friedjung M. 2000, \aaps, 146, 407 

\bibitem[1980]{cant80} Canto, J., Meaburn, J., Theokas, A.~C., Elliott, K.~H. 
1980, MNRAS, 193, 911

\bibitem[1994]{chav94} Chavarria-K. C., Moreno-Corral, M.~A.,
  Hernandez-Toledo, H., Terranegra, L., de Lara, E., 1994, \aap, 283, 963

\bibitem[1998]{corc98} Corcoran, M., Ray, T.~P. 1998, \aap, 336, 535

\bibitem[2010]{corr10} Corradi, R.~L.~M., Giammanco, C. 2010, 520, A99 

\bibitem[2008]{corr08} Corradi, R.~L.~M., Rodr{\'{\i}}guez-Flores, E.~R., 
Mampaso, A. et al. 2008, \aap, 480, 409

\bibitem[1985]{fink85} Finkenzeller, U. 1985, \aap, 151, 340

\bibitem[1992]{hama92} Hamann, F., Persson, S.~E. 1992, ApJS, 82, 285

\bibitem[1994]{hart94} Hartigan, P., Morse, J.~A., Raymond, J. 1994, ApJ, 436, 125

\bibitem[1960]{herb60} Herbig, G.~H., 1960, ApJS, 4, 337

\bibitem[2004]{hern04} Hern{\'a}ndez, J., Calvet, N., Brice{\~n}o, C., Hartmann,
  L., Berlind, P. 2004, AJ, 127, 1682

\bibitem[1992]{hill92} Hillenbrand, L.~A., Strom, S.~E., Vrba, F.~J., Keene, J. 1992, ApJ,
  397, 613

\bibitem[2008]{iked08} Ikeda, H., Sugitani, K., Watanabe, M., Fukuda, N. et al. 2008, AJ, 135, 2323

\bibitem[1997]{koho97} Kohoutek, L., Wehmeyer, R., 1997, 
Catalogue of stars in the northern Milky Way having H-alpha in emission (Hamburg: Sternwarte) 

\bibitem[1995]{mass95} Massey, P., Johnson, K.~E., Degioia-Eastwood, K., 1995, ApJ, 454, 151

\bibitem[2008]{math08} Mathew, B., Subramaniam, A., Bhatt, B.~C. 2008, MNRAS, 388, 1879

\bibitem[2010]{math10} Mathew, B., Subramaniam, A., Bhavya, B. 2010, Bulletin of the Astronomical 
Society of India, 38, 35

\bibitem[2007]{negu07} Negueruela, I., Marco, A., Israel, G.~L., Bernabeu, G. 2007, \aap, 471, 485

\bibitem[2006]{oste06} Osterbrock, D.~E., Ferland, G.~J. 2006, Astrophysics of
  gaseous nebulae and active galactic nuclei (2nd ed.:~Sausalito, CA: University Science Books)

\bibitem[1985]{riek85} Rieke, G.~H., Lebofsky, M.~J. 1985, ApJ, 288, 618

\bibitem[1985]{royj85} Roy, J.-R., Joncas, G. 1985, ApJ, 288, 142

\bibitem[1989]{schm89} Schmid, H.~M. 1989, \aap, 211L, 31S

\bibitem[1982]{schm82} Schmidt-Kaler, Th. 1982, 
Landolt-B\"ornstein: Numerical Data and Functional Relationships in Science and 
Technology, ed. K. Schaifers, \& H.H. Voigt (Berlin: Springer), VI/2b

\bibitem[1991]{sugi91} Sugitani, K., Fukui, Y., Ogura, K. 1991, ApJS, 77, 59

\bibitem[1998]{wate98} Waters, L.~B.~F.~M., Waelkens, C. 1998, \araa, 36, 233

\bibitem[1975]{webs75} Webster, B.~L., Allen, D.~A. 1975, MNRAS, 171, 171

\end{thebibliography}
\end{document}